# MODELING PATIENT FLOW IN THE EMERGENCY DEPARTMENT USING MACHINE LEARNING AND SIMULATION

**Emad Alenany**
State University of New York at Binghamton, USA
ealenan1@binghamton.edu

**Abdessamad Ait El Cadi**
LAMIH UMR CNRS 8201, Université Polytechnique Hauts-de-France (UPHF), Valenciennes, France
Abdessamad.AitElCadi@uphf.fr

**ABSTRACT:** *Recently, the combination of machine learning (ML) and simulation is gaining a lot of attention. This paper presents a novel application of ML within the simulation to improve patient flow within an emergency department (ED). An ML model used within a real ED simulation model to quantify the effect of detouring a patient out of the ED on the length of stay (LOS) and door-to-doctor time (DTDT) as a response to the prediction of patient admission to the hospital from the ED. The ML model trained using a set of six features including the patient age, arrival day, arrival hour of the day, and the triage level. The prediction model used a decision tree (DT) model, which is trained using historical data achieves a 75% accuracy. The set of rules extracted from the DT are coded within the simulation model. Given a certain probability of free inpatient beds, the predicted admitted patient is then pulled out from the ED to inpatient units to alleviate the crowding within the ED. The used policy combined with adding specific ED resources achieve 9.39% and 8.18% reduction in LOS and DTDT, respectively.*

**KEYWORDS:** *Healthcare systems, Emergency department, Simulation, Machine learning techniques, management.*

## 1 INTRODUCTION

The hospital system is a network of interconnected units: ED, intensive care unit (ICU), operating rooms (OR), and inpatient nursing units (NU) (Kolker, 2013). EDs face significant challenges regarding patient delay and safety (Yarmohammadian et al., 2017). A significant portion of admission to different hospital inpatient units (IU) is coming from the ED (Morganti et al., 2013). Given the scarce ED resources, a significant amount of research has been done to investigate methods to decrease the length of stay and streamline flows at EDs.

ED overcrowding could lead to adverse results on patient safety (Cameron, 2006). Therefore, many mechanisms have been developed to mitigate the ED overcrowding as studying different process changes and their effects through simulation (e.g., Hussein et al., 2017). Yarmohammadian et al. (2017) provide a review of different strategies to deal with ED overcrowding. The strategies adopted include patient streaming, fast track, and team triage, among others.

Ghanes et al. (2015) present a simulation optimization for resource staffing in the ED. The objective is to minimize the average LOS subject to the limit on the DTDT for most severe patients and budget constraints. The authors performed a sensitivity analysis and showed the tradeoff between LOS and DTDT. Yousefi et al. (2018) provide resource allocation optimization for an ED by developing a meta-model using an ensemble of adaptive network-based fuzzy neuro inference systems, feed-forward neural network, and recurrent neural network. Then, they used a genetic algorithm to optimize the objective function generated from the metamodel. Rabbani et al. (2018) Solve the simulation optimization resource allocation problem for an integrated model of an ED with pharmacy, and lab using artificial neural networks and data envelopment analysis.

The use of machine learning in conjunction with simulation models is gaining a lot of attention in the last few years (Greasley and Edwards, 2019). Greasley and Edwards (2019) present a recent review for studies of combining simulation with big data analytics. Greasley (2019) describes ways of combining simulation with ML. Lee et al. (2015) is considered one of the first studies to combine ML, simulation, and optimization to improve ED operations. In their study, the authors utilized patient characteristics, demographics, hospital operations, socioeconomic status, clinical information, and disease behavioral patterns to predict patient ED readmission. Gartner and Padman (2020) studied the factors associated with the perception of waiting time, i.e., under, over, or correct estimation of the waiting times. Then, they used ranking and selection methods to identify the importance of such features. Through using different machine learning classification algorithms, they have achieved 70% and 78% for correctly classifying a specific patient to be overestimating his waiting time in the waiting area and the treatment room, respectively. The authors then fed a simulation model with the results from the classification model to test the effects of different staff changes on patient satisfaction.

With ML and data mining tools become widely used, researchers have studied the potential of those tools to predict patient admission from the ED toward better management of patient flow in the ED. A recent review of ML applications in the ED is given in (Shafaf and Malek, 2019). Cameron et al. (2015) present a logistic regression model to anticipate the likelihood of admission at the triage point. They used two years of regularly collected data from hospitals in Galacso. The authors found the highest important indicators are "triage classification, age, National Early Warning Score, appearance by rescue vehicle, referral source, and admission within the last year" (Cameron et al., 2015). They achieve an area under the curve of the operating characteristic curve (AUC-ROC) of 0.877. Graham et al. (2019) provided a comparison between three ML models to predict patient admission from ED: logistic regression, decision tree, and gradient boosting machines. The authors achieve an accuracy of 79.94% and an AUC-ROC of 0.849. Liu et al. (2019) used a deep belief network (DBN), k-nearest neighbor (kNN), and artificial neural network for classification of ED patients attendance/disposal from the ED. They have identified 5 different classes of patients: inpatient admission, discharged with follow-up needed, fracture clinic referral, outpatient clinic referral, and transfer to another healthcare entity. They achieved 76.20% accuracy using DBN.

It is shown from previous studies they provided stand-alone studies for prediction models for patient admission to the hospital from the ED. The current study contributes to the literature of integrating ML with simulation by utilizing a prediction model within the simulation model to manage the patient flow in the ED. To the best of our knowledge, no available study has been performed that quantify the effect prediction models utilization within simulation for early diversion of predicted admitted patients from ED to IU on the LOS and DTDT.

The remainder of this paper goes as follows. Section 2 presents a system description for the ED under consideration, while section 3 presents the methodology including simulation and prediction models. Results and discussions are shown in section 4. Finally, section 5 presents conclusions and future work.

## 2 SYSTEM DESCRIPTION

The ED system considered here is recently studied in (Cheaitou et al., 2020). The patient flow at the ED is shown in Figure 1. A typical patient trajectory starts with the triage process where the arriving patient goes for the triage by a triage nurse. After triage, patients are categorized into one of five severity levels (Ward, 2006). The most severe is level 1, which they need prompt care, while level 5 is the least critical. Critical patients (level 1) goes on for first aid. Then, all patients do diagnostic tests (radiology and/or lab tests) if necessary. Next, needed patients pass by a complementary treatment step, and later they are either discharged to home, admitted to hospital inpatient units, or transferred to another hospital.

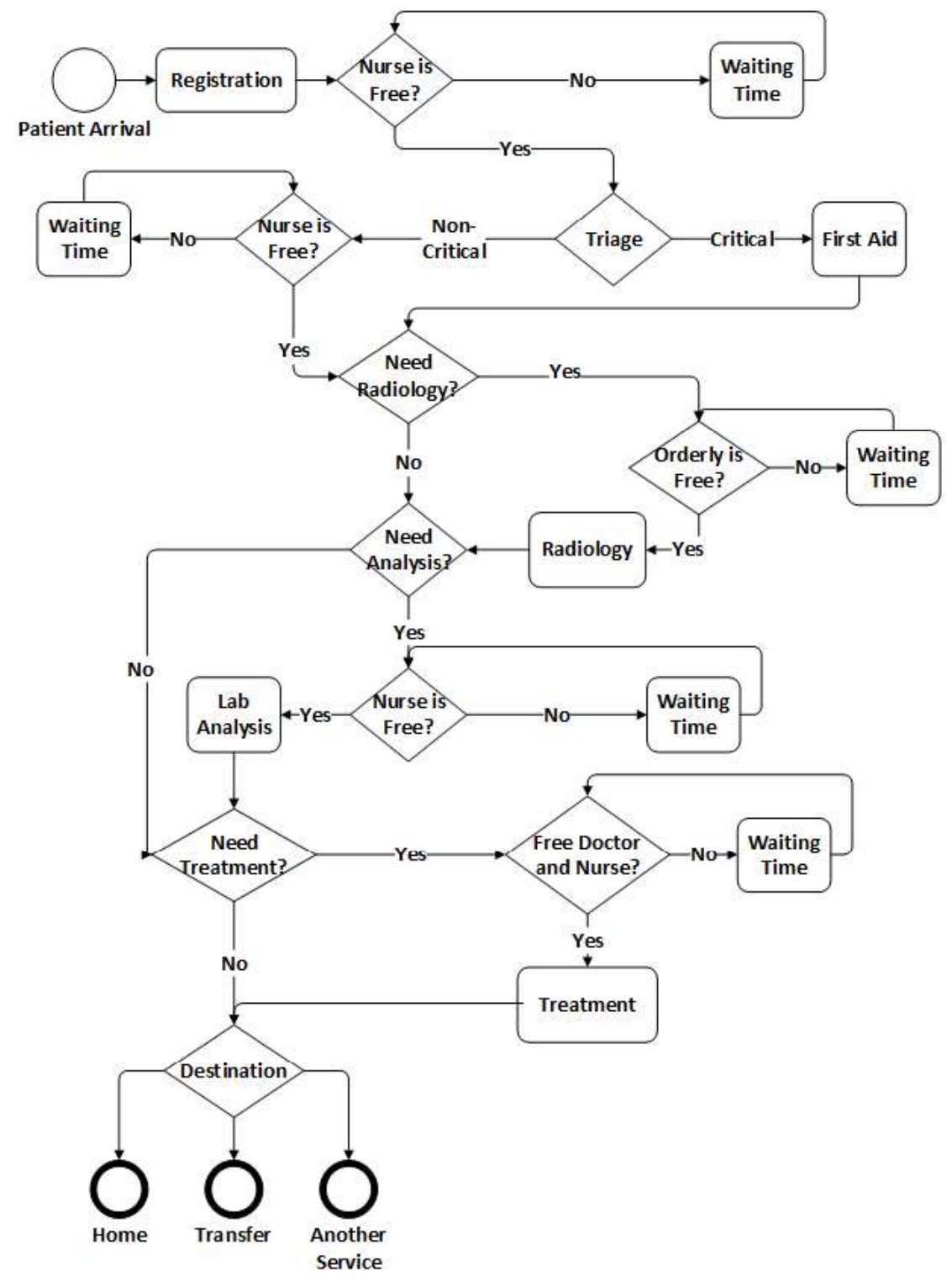

Figure 1 patient flow in the ED system

The ED system works for two shifts with different resource schedules. The first shift from 08:00 am to 04:00 pm with three physicians (two Emergency specialists + one Resident), seven Registered Nurses, two Orderlies, and two Receptionists. The second shift starts at 04:00 pm to 08:00 am and has four physicians (one Emergency specialist + three Residents), seven Registered Nurses, two Orderlies, and two Receptionists. For the radiology technicians, they work on two shifts: 7:00 am-7:00 pm (4 technicians), and 7:00 pm to 7:00 am (two technicians).

## 3 METHODOLOGY

This section describes the simulation and prediction models and their constituents.

### 3.1 Simulation model

The probability distributions used in the simulation model for the arrival and different service processes are shown in table 1. They are fitted from the historical records data and experts opinion (Cheaitou et al., 2020). The simulation is developed using Rockwell Arena simulator V15. The simulation run length is set to 1 year with 5 days as a warm-up period and 10 replications. The LOS is considered the main key performance indicator for the ED system in this study. The average LOS from the simulation is found to be 98.68 minutes, while the average actual LOS is 100 minutes, which seems reasonable to validate the model.

| Parameters | Distribution functions of random variables (minutes) |
|---|---|
| Arrival rate | Exponential(24) |
| Registration time | Uniform(3,10) |
| Triage time | Triangular(5, 10, 15) |
| First Aid for PICU cases | Uniform(10,45) |
| First Aid for ICU cases | Uniform(20,60) |
| First Aid for CCU cases | Uniform(30,90) |
| Complementary tests durations: laboratory | Triangular(15, 45, 90) |
| Complementary tests durations: radiology | Triangular(15, 45, 90) |
| Complementary Treatment | Uniform(10,60) |

Table 1: Simulation parameters

Using the validated simulation model, a few scenarios of adding resources are tested to improve patient LOS. There are three tested scenarios. The baseline scenario reflects the simulation model of the original ED system. Other scenarios include adding one more nurse to reduce the waiting times during the treatment phase (scenario A). Another is adding one more nurse in addition to one more orderly (Scenario B). The motivation for Scenario B is there has been a lack of orderlies required for the radiological examination. The results are given in section 4.

### 3.2 Predicting hospital admission from the ED within simulation

The successful prediction of ED patients as becoming inpatients could help to better patient flow management. Also, it could help reduce ED congestion by avoiding the excessive use of scarce ED resources.

The current subsection uses ML to predict patient hospital admission from the ED. The prediction model is trained on generated data using DT and kNN algorithms. Then, the trained prediction model is implemented within the simulation model to predict the patient admission status given the specific set of patient characteristics: admission status, gender, age, arrival day, arrival hour, triage level, and having X-Ray or Lab tests or not. If the patient is predicted to move to an IU, the patient will be directed to IU and removed from the ED given available beds at the IU. Figure 4 shows the flowchart with the proposed changes to the simulation model.

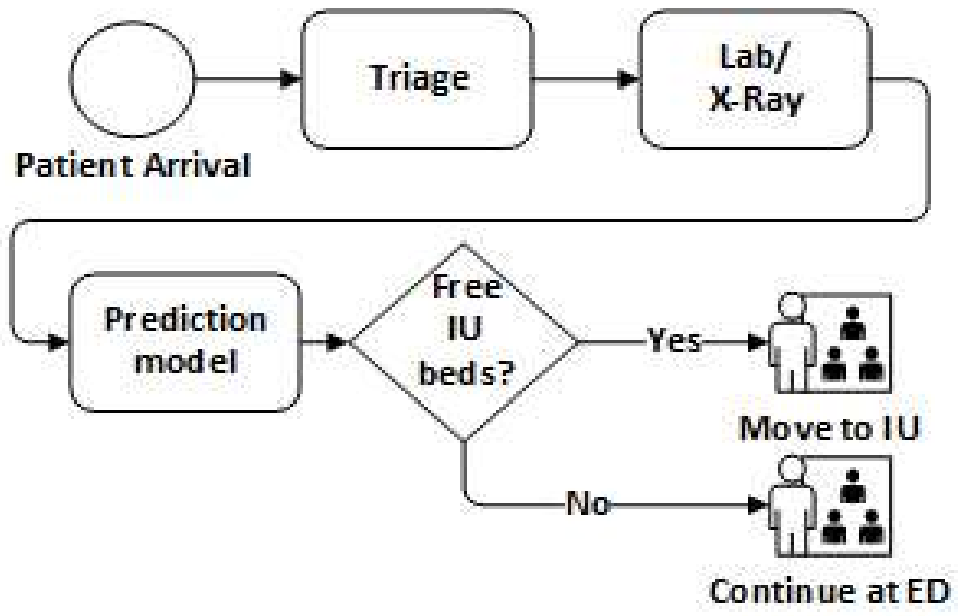

Figure 4 Integration of simulation and prediction model

The prediction model methodology includes data generation, training, and testing. Here DT and kNN are selected as the prediction model for their common use in previous studies for the same problem. They were also found to achieve effective results compared to other tested models. The trained model is integrated into the simulation. This is done by translating the DT rules into Visual Basic for Applications (VBA) code within the Arena simulation model.

The data generated for the different features are inspired by the summary statistics given in recent studies (e.g., Graham et al., 2018). A number of 500 patient records with six features are generated. Table 2 shows summary statistics for generated patient records to the ED used in the prediction model construction. The rate of admitted/non-admitted patients of the generated data is like other studies (e.g., Graham et al., 2018 and references therein).

The accuracy, specificity, and sensitivity are used to assess the prediction model performance. Given the information about the actual and predicted status for patients, these measures calculated as follows (Baratloo et al., 2015):

$$Accuracy = \frac{TP + TN}{TP + TN + FP + FN} \quad (1)$$

$$Sensitivity = \frac{TP}{TP + FN} \quad (2)$$

$$Specificity = \frac{TN}{TN + FP} \quad (3)$$

where

True-positive (TP) = the number of times a patient correctly predicted as admitted

False-positive (FP) = the number of times a patient incorrectly predicted as admitted

True-negative (TN) = the number of times a patient correctly predicted as not admitted

False-negative (FN) = the number of times a patient incorrectly predicted as not admitted

The data is then divided into training and testing with 70% and 30%, respectively. The prediction model is done using 'tree' and 'class' libraries in R for DT and kNN models, respectively. Table 3 shows the performance of different prediction models using DT and kNN. The first DT model (DT1) used all the patient features including X-Ray/Lab test. It achieves high accuracy and specificity of 0.81 and 0.89, respectively. However, sensitivity is only 35%. The second DT (DT2) model excludes the X-Ray/Lab test feature. DT2 has the incentive to predict for the patient admission status from an earlier point, which is directly after the point of triage. This model achieves an accuracy of 0.75. Using a kNN model with only two numeric features, age, and arrival hour of the day, with the number of neighbors, k = 1, the achieved accuracy is 0.74, and sensitivity 0.40. The high difference between specificity and sensitivity of the models used could be attributed to the imbalance between admitted/non admitted records in the dataset (Graham et al., 2018). Increasing the sample size and using specific methods to handle data imbalance would be tested in a future study.

As a validity to the reported performance of the DT model of this study, the last row of table 4 shows the reported performance from a recent study (Graham et al., 2018). Graham et al. (2018) reported 0.80, 0.90, and 0.53 for the accuracy, specificity, and sensitivity, respectively. The results seem close to the DT used in this study, with a higher sensitivity result.

## 4 RESULTS AND DISCUSSION

Figure 5 shows the decision tree for the DT2 model. The triage level is included in the input variables, however, it is not included in the construction of the tree by the decision tree algorithm. The DT prediction models used only four of the input features: age, registration date, registration time, and gender. The first branch in DT 2 splits the tree based on the patient Age, which indicates that Age is the most important predictor for patient admission status. This result agrees with reported results of other studies (e.g., Sun et al., 2011).

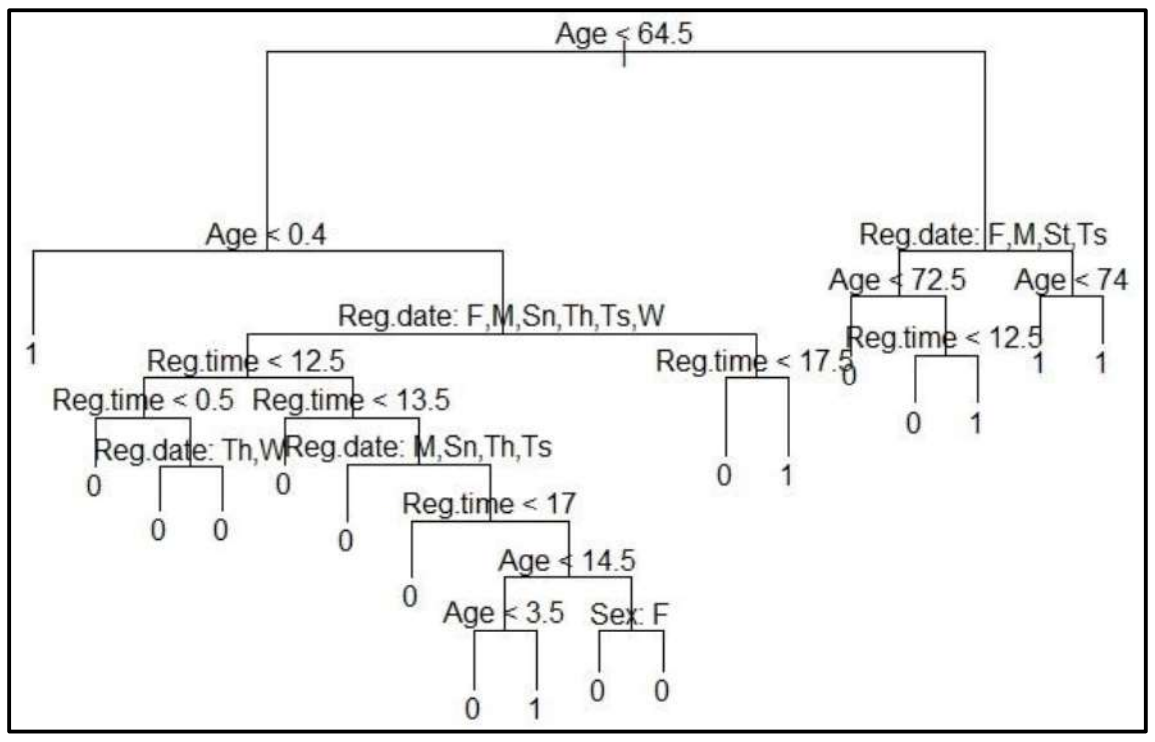

Figure 5 Tree plot of the DT2 model

| Feature | Main Categories | % per category |
|---|---|---|
| Admission | Yes<br>No | 20.1<br>79.9 |
| Gender | Female<br>Male | 51.9<br>48.1 |
| Triage | L3<br>L4 | 49.4<br>48.2 |
| X-Ray | Yes<br>No | 54.4<br>45.6 |
| Lab | Yes<br>No | 52.2<br>47.8 |
| Age | Mean<br>SD | 31.7<br>24.4 |
| Arrival day | Friday<br>Sunday<br>Tuesday<br>Wednesday | 21.4<br>17.7<br>17.5<br>20.9 |
| Arrival hour | 4 pm<br>5 pm<br>6 pm<br>7 pm<br>9 pm<br>10 pm | 6.7<br>6.3<br>7.7<br>5.8<br>6.7<br>6.3 |

Table 2 Summary statistics of generated patients records

| | Accuracy | Specificity | Sensitivity |
|---|---|---|---|
| DT1 | 0.81 | 0.89 | 0.35 |
| DT2 | 0.75 | 0.88 | 0.26 |
| kNN (k=1) | 0.74 | 0.79 | 0.40 |
| Graham et al. (2018) DT | 0.80 | 0.90 | 0.53 |

Table 3 Summary of the prediction models performance

The following set of rules are extracted from DT2 model to predict admission a specific patient:

1. Patients with $Age < 0.4$ years old.
2. Patients with $Age < 64.5$ and $Age > 0.4$ years old who arrive after 5:30 pm on Saturday.
3. Patients with $Age > 3.5$ and $Age < 64.5$ years old who arrive after 5:00 pm on Wednesday or Friday.

4. Patients with $Age > 64.5$ and $Age < 74$ who arrive on Sunday, Wednesday, or Thursday.
5. Patients with $Age > 72.5$ and arrive on Friday, Saturday, Monday, or Tuesday and arrive at 12:30 pm or later during the day.

The rules are coded with VBA within the simulation model. Table 4 presents the simulation results for the three scenarios described in the previous section, in addition to the inclusion of ML with each scenario. The average LOS is presented, along with the DTDT. The DTDT time represents the time to first treatment by a doctor in the ED for critical patients (Ghanes et al., 2015).

Using the "Baseline+ML" scenario improves the LOS and average DTDT by 3.66% and 1.61%, respectively. Using "scenario A", LOS and DTDT improves by 5.46% and 7.49%, respectively. However, combining ML with "scenario A" (scenario A+ML) leads to higher LOS and DTDT reduction compared to "scenario A". The LOS and DTDT could be reduced by 8.53% and 8.14% relative to the baseline scenario in "scenario A+ML". Similar results could be obtained in "scenario B+ML", where expected improvement to the patient logistical flow in the ED from adding more resources is enhanced using the ML model. All scenarios are statistically significant relative to the baseline scenario.

## 5 CONCLUSIONS

This study provided a novel application of combining ML with a simulation model for a real ED to model the ED patient flow. ML is used in the form of a prediction model to anticipate whether the ED patient would be finally admitted to the inpatient unit after going through treatment at the ED. With the output from the ML model, the simulation model is used to assess the expected reduction in patient LOS and DTDT if those patients are admitted directly to inpatient units at an early stage of their ED journey, given an available inpatient bed.

A decision tree model is trained to predict whether an arriving patient will be admitted to the hospital given a specific set of patient data includes age, registration date, registration time, and gender. The extracted rules from the trained model are then coded into the simulation model. The results show that the expected decrease in LOS and DTDT could lead to a reduction of more than 9% and 8% in the patient LOS and DTDT, respectively.

This research shows the potential of using ML and simulation for better ED patient flow management to alleviate ED congestion. Future work would include increasing the data size used for the ML model training. Also, handling the data imbalance in the class variable (Admitted vs. Not admitted). Other ML models could be considered to enhance the performance as well. This study focused on LOS and DTDT measures due to the direct implementation in the current simulation model. The effect of other related measures on patient quality should be considered as well.

| Scenario | LOS (minutes, % decrease) | DTDT (minutes, % decrease) |
|---|---|---|
| Baseline | 98.68 | 19.04 |
| Baseline+ ML | 95.64* (-3.66%) | 18.73* (-1.61%) |
| Scenario A | 93.29* (-5.46%) | 17.62* (-7.49%) |
| Scenario A+ML | 90.26* (-8.53%) | 17.49* (-8.14%) |
| Scenario B | 92.15* (-6.62%) | 17.62* (-7.44%) |
| Scenario B+ML | 89.41* (-9.39%) | 17.48* (-8.18%) |

Table 4 Simulation results of the baseline, scenario A, and scenario B with/without the ML model. * t-test significant at $p < 0.05$